\documentclass[aps,prx,reprint,superscriptaddress,noeprint,longbibliography]{revtex4-2}
\usepackage{siunitx}
\usepackage{graphicx}			

\usepackage{amsmath}
\usepackage{amsfonts}
\usepackage{amssymb}
\usepackage{color}
\usepackage{xcolor}
\usepackage{bbold}

\usepackage{color}
\usepackage{orcidlink}

\begin{document}
\newcommand{\AR}[1]{{\color{red} AR: #1}}
\renewcommand{\vec}[1]{{\boldsymbol #1}}
\newcommand{\bsts}{BiSbTeSe$_2$}     
\newcommand{\bsb}{Bi$_{1.5}$Sb$_{0.5}$Te$_{1.7}$Se$_{1.3}$}
\newcommand{\bsn}{Bi$_{1.08}$Sn$_{0.02}$Sb$_{0.9}$Te$_{2}$S}


\title{Microscopic origin of an exceptionally large phonon thermal Hall effect from charge puddles in a topological insulator}


\author{Rohit Sharma\,\orcidlink{0000-0001-9815-0733}}
\altaffiliation{Present address: Institute for Topological Insulators, Universit\"at W\"urzburg, Am Hubland, 97074 W\"urzburg, Germany}
\affiliation{II. Physikalisches Institut, Universit\"at zu K\"oln, Z\"ulpicher Str.\ 77, 50937 K\"oln, Germany}
\author{Yongjian Wang\,\orcidlink{0000-0001-5921-1986}}
\altaffiliation{Present address: High Magnetic Field Lab. of Anhui Province, HFIPS, Chinese Academy of Sciences, Hefei 230031, China}
\affiliation{II. Physikalisches Institut, Universit\"at zu K\"oln, Z\"ulpicher Str.\ 77, 50937 K\"oln, Germany}
\author{Yoichi Ando\,\orcidlink{0000-0002-3553-3355}}
\affiliation{II. Physikalisches Institut, Universit\"at zu K\"oln, Z\"ulpicher Str.\ 77, 50937 K\"oln, Germany}
\author{Achim Rosch\,\orcidlink{0000-0002-6586-5721}}
\email{rosch@thp.uni-koeln.de}
\affiliation{Institute for Theoretical Physics, Universit\"at zu K\"oln, Z\"ulpicher Str.\ 77, 50937 K\"oln, Germany}
\author{Thomas Lorenz\,\orcidlink{0000-0003-4832-5157}}
\email{tl@ph2.uni-koeln.de}
\affiliation{II. Physikalisches Institut, Universit\"at zu K\"oln, Z\"ulpicher Str.\ 77, 50937 K\"oln, Germany}

\date{\today}

\begin{abstract}
 We present the experimental observation of a drastically enhanced thermal Hall effect in the topological insulator material TlBi$_{0.15}$Sb$_{0.85}$Te$_2$. Although heat transport is dominated by phonons, moderate magnetic fields generate a thermal Hall ratio ($\kappa_{xy}/\kappa_{xx}$) above 2\%, an unprecedented value for a nonmagnetic material. The transverse thermal conductivity $\kappa_{xy}$ exhibits  a pronounced maximum in fields of a few Tesla. This characteristic field dependence allows us to identify the microscopic origin of the thermal Hall effect in this system. Small densities of charged impurities induce locally conducting regions, so-called charge puddles, within the bulk insulating matrix. Via electron-phonon coupling, these charge puddles imprint a large thermal Hall effect onto the phonons accounting for both the magnitude and the magnetic-field dependence of the observed effect.
\end{abstract}


\maketitle

\section{Introduction}

The thermal Hall effect (THE) has attracted considerable scientific interest due to its unique ability to detect responses from both mobile charge carriers as well as charge-neutral quasi particles, such as phonons or magnons \cite{PhysRevLett.95.155901, 2010Sci...329..297O}. In simple metals the main carriers of heat are electrons or holes. Here, in the presence of a magnetic field $B$, the charged quasi particles are deflected by the Lorentz force giving rise to a Hall ratio $\kappa_{xy}/\kappa_{xx}$ of the order of $\omega_c \tau$, where $\omega_c$ is the cyclotron frequency and $\tau$ the electronic scattering time.

Such a Lorentz-force induced THE cannot be present in insulators with charge-neutral quasi particles. In magnetic insulators a finite THE has been assigned to magnons, originating, for example, from a Berry curvature as a consequence of a nontrivial topology of magnon bands, facilitated by the Dzyaloshinskii-Moriya interaction in specific lattice structures like pyrochlore, honeycomb, kagome, triangular \cite{PhysRevB.85.134411, 2015Sci...348..106H, 2010Sci...329..297O, PhysRevLett.115.106603, 2021PhRvL.127x7202Z, 2024NatCo..15..243K}. Remarkably, phonons can also give rise to a THE, as was experimentally observed for the first time in the paramagnetic insulator Tb$_3$Ga$_5$O$_{12}$ \cite{PhysRevLett.95.155901}, and was theoretically discussed in the context of Raman-type interactions between large spins and phonons, the Berry curvature of phonon bands, and skew scattering of phonons by superstoichiometric Tb$^{3+}$ ions \cite{PhysRevLett.124.186602,PhysRevLett.96.155901,PhysRevLett.105.225901,PhysRevB.86.104305,PhysRevLett.113.265901}. Over the last years, experimental evidence of sizable phonon-induced THE has been observed in a diverse set of different materials, ranging from multiferroics  like Fe$_2$Mo$_3$O$_8$ \cite{2017NatMa..16..797I} to cuprate superconductors \cite{2019Natur.571..376G, 2020NatPh..16.1108G}, phonon glass systems like Ba$_3$CuSb$_2$O$_9$ \cite{PhysRevLett.118.145902}, Mott insulators \cite{2020NatCo..11.5325B}, metallic spin ice materials Pr$_2$Ir$_2$O$_7$ \cite{2022NatCo..13.4604U}, the quantum spin liquid RuCl$_3$ \cite{PhysRevX.12.021025} and the antiferromagnetic insulator Cu$_3$TeO$_6$ \cite{2022PNAS..11908016C}. In non-magnetic SrTiO$_3$, a phonon THE has been linked with the presence of antiferrodistortive structural domains~\cite{PhysRevLett.124.105901}, but in isotopically $^{18}$O substituted and Ca-doped samples the THE is substantially suppressed, which could be related with the stabilization of ferroelectric order \cite{PhysRevLett.126.015901, 2022PNAS..11901975J, PhysRevB.100.195121}. A phonon-induced THE has even been  detected in elemental black phosphorus~\cite{Black-Phos_Li_2023NatCo..14.1027L}, Si, Ge, and trivial  non-magnetic insulators like MgO, MgAl$_2$O$_4$~\cite{Universal_phonon-THE_Jin-PRL2025}, or  Y$_2$Ti$_7$O$_7$~\cite{PhysRevB.110.L100301}. Thus, from the experimental perspective, a sizable phonon Hall effect with Hall ratios $\kappa_{xy}/\kappa_{xx}$ of the order  of $10^{-3}$ in fields of order 10\,T, sometimes reaching values up to $10^{-2}$, appears to be ubiquitous across a wide range of material systems.

Several theoretical models have been proposed to understand a phonon-induced THE, which can be categorized into intrinsic and extrinsic scenarios. Intrinsic mechanisms encompass factors like Berry curvature of phonon bands \cite{PhysRevLett.105.225901,PhysRevB.86.104305,PhysRevX.8.031032}, intrinsic phonon scattering by collective fluctuations \cite{PhysRevB.106.245139}, and interactions between phonons and other quasi-particles like magnons or spinons \cite{PhysRevB.104.035103, PhysRevLett.123.167202,PhysRevLett.121.147201,PhysRevX.8.031032}. Conversely, extrinsic factors explore phonon scattering by impurities or defects shaping THE \cite{PhysRevLett.124.167601,PhysRevB.103.205115,PhysRevB.106.144111,2022PNAS..11915141G, PhysRevB.105.L220301}. Despite a strong surge in research interest from both theoretical and experimental perspectives, there is still no consensus on the origin or mechanism of the phonon THE. In particular, it remains challenging to explain the widespread occurrence and the size of the THE and to find a way to pinpoint its physical origin.

Here, we report a huge THE in the weakly charge-carrier compensated 3D Topological Insulator (TI) TlBi$_{0.15}$Sb$_{0.85}$Te$_2$, which substantially differs from the above described previous THE data and allows us to identify the underlying microscopic mechanism. First of all, this concerns the absolute value of the Hall ratio $\kappa_{xy}/\kappa_{xx}$ that lies in the range of 2\%  for moderate magnetic fields of 2 to 8\,T over a wide temperature range from about 50 to 150\,K and, secondly,  the magnetic-field dependence $\kappa_{xy}(B)$ is strongly non-monotonic. In contrast, the phonon-related THE in the above mentioned materials typically show field-linear $\kappa_{xy}(B)$ with maximum Hall ratios in the $10^{-3}$ range observed at specific temperatures close to the characteristic maxima of the phonon-dominated longitudinal $\kappa_{xx}(T)$ \cite{2022PNAS..11915141G,Black-Phos_Li_2023NatCo..14.1027L}.

\begin{figure}[tbp]
	\centering
    \includegraphics[width=0.97\columnwidth]{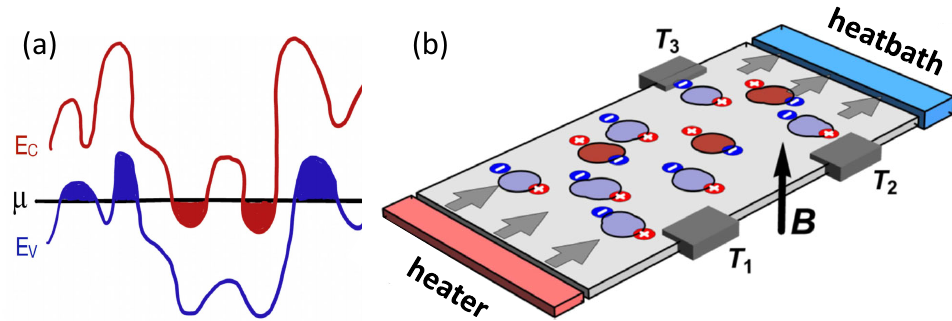}
	{\caption{(a) Spatially varying energy spectrum in a compensated TI containing charged impurities \cite{ShklovskiiEfros1972,PhysRevLett.109.176801}. The lines represent the conduction and valence band edges, E$_C$ and E$_V$, respectively, and the shaded areas indicate the formation of electron- or hole-type charge puddles in regions where the chemical potential $\mu$ crosses either  E$_C$ or E$_V$. (b)  Schematic setup to measure the longitudinal and transverse temperature differences ($\Delta\textrm{T}_\textrm{x} = T_1-T_2$,  $\Delta\textrm{T}_\textrm{y} = T_2-T_3$). A longitudinal heat current from the heater to the heat bath is mainly carried by phonons inducing temperature gradients in electron and hole puddles. Due to the electronic $\kappa^{ch}_{xy}$, the puddles develop a temperature gradient in perpendicular direction (indicated by $+$ and $-$ in the red and blue dots), heating the phonons. This results in a transversal temperature gradient of the phonon system dominantly from hole puddles (see text). }
		\label{FIG.schematic}}
\end{figure}
We explain the anomalous THE in TlBi$_{0.15}$Sb$_{0.85}$Te$_2$ by a novel mechanism that arises from the presence of  tiny densities (a few ppm) of charged impurities in a material with a large dielectric constant, $\epsilon\approx 10^2$~\cite{Borgwardt_PhysRevB.93.245149,epsilon_TlSbT2_Deger2015}. Such defects occur unavoidably during growth and to obtain bulk-insulating compounds, the density of positively and negatively charged defects have approximately to be the same. They change, however, profoundly the bulk properties of three-dimensional materials as pointed out early by Shklovskii and Efros \cite{ShklovskiiEfros1972}. Coulomb charges $Z_i=\pm 1$ placed at random positions $\vec R_i$ result in a Coulomb potential $V_0(\vec r)=\sum_i \frac{Z_i e^2}{4 \pi \epsilon_0 \epsilon  |\vec r-\vec R_i|}$. 
In a region of radius $R$, the potential fluctuations grow with $\sqrt R$, reaching a size of the order of the bulk gap $\Delta$ at a large length scale $R_g \approx d_\text{def} (\Delta/E_c)^2 /\ln(\Delta/E_c)$   much larger \cite{PhysRevLett.109.176801,Borgwardt_PhysRevB.93.245149,2017PhRvB..96g5204B} than the distance of defects $d_\text{def}$, as the typical Coulomb energy  $E_c=\frac{e^2}{4 \pi \epsilon_0 \epsilon d_\text{def}}$ is much smaller than the gap, $E_c/\Delta \sim 10^{-2}$. As a consequence, in a highly non-linear and non-local screening process \cite{ShklovskiiEfros1972,2017PhRvB..96g5204B} so-called charge puddles form, locally conducting regions with electron or hole doping embedded in an insulating matrix, see Fig.~\ref{FIG.schematic}. 
The mobile charges in the  puddles can directly be observed in optical conductivity  \cite{Borgwardt_PhysRevB.93.245149,PhysRevB.99.161121} and STM experiments \cite{Borgwardt_PhysRevB.93.245149,Beidenkopf2011} and have a profound effect on a wide range of electric transport properties~\cite{PhysRevLett.109.176801,Borgwardt_PhysRevB.93.245149,PhysRevB.96.195135,PhysRevB.99.161121,Trang2016,Breunig_2017NatCom_815545B}. In systems like TlBi$_{0.15}$Sb$_{0.85}$Te$_2$ where $\Delta/E_c\sim 10^2$, these puddles are especially large but charged impurities necessarily create dynamical electronic defects in a much broader class of systems.

We will argue that the Lorentz force acting on mobile charge carriers within localized charge puddles imprints a large THE onto the phonons, as sketched in Fig.~\ref{FIG.schematic}(b).

\begin{figure}[!tbp]
	\centering
	\hspace*{-0.6cm}
	\includegraphics[width=\columnwidth]{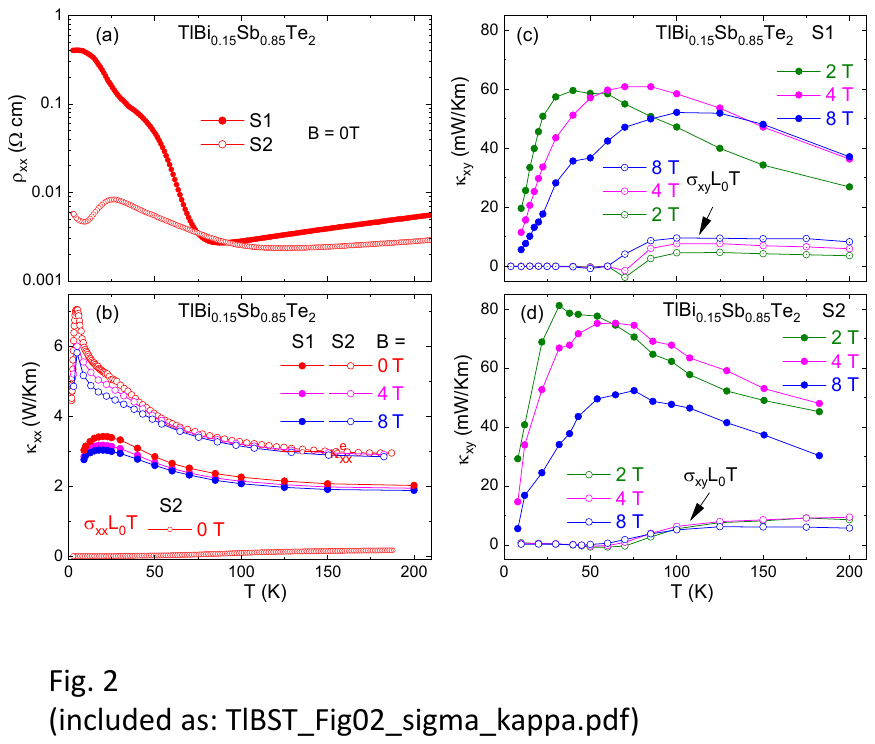}
	\caption{(a) Electrical conductivities $\rho_{xx}$ and (b) thermal conductivities $\kappa_{xx}$ of TlBi$_{0.15}$Sb$_{0.85}$Te$_2$ samples S1 and S2. In (b), we also show the estimated contribution $\sigma_{xx}L_0T$ from mobile charge carriers based on the Wiedemann-Franz law for the better conducting sample S2.    
	(c,d) Thermal Hall conductivities $\kappa_{xy}$ of both samples for different magnetic fields, which strongly exceed the estimated charge-carrier contributions $\sigma_{xy}L_0T$  based on the electrical Hall conductivities.}
	\label{Fig_rho_kappa}
\end{figure}

\section{Transport experiments}
Figure~\ref{Fig_rho_kappa}(a) shows the temperature-dependent longitudinal electrical resistivities $\rho_{xx}$ of two TlBi$_{0.15}$Sb$_{0.85}$Te$_2$ samples from different batches of the as-grown crystals. Both samples display a metallic $\rho_{xx}$ at high temperatures, undergoing a metal-to-insulator-like transition around 80\,K for sample S1, while in S2 a slight semiconducting behavior evolves below 100~K followed by additional anomalies below 30~K. The different  low-temperature behavior of $\rho_{xx}$  can be attributed to a varying degree of compensation in different batches of TlBi$_{0.15}$Sb$_{0.85}$Te$_2$, leading to different degrees of disorder. The absolute values of $\rho_{xx}$ for both samples are
lower than previously reported for other TlBi$_{0.15}$Sb$_{0.85}$Te$_2$ samples of nominally the same composition~\cite{Breunig_2017NatCom_815545B} and for related TI materials of the BiSbTeSe$_2$ type~\cite{Borgwardt_PhysRevB.93.245149,PhysRevB.99.161121}, indicating that positively and negatively charged impurities are not perfectly compensated in these TlBi$_{0.15}$Sb$_{0.85}$Te$_2$ samples. In Fig.~\ref{Fig_rho_kappa}(b) the thermal conductivities $\kappa_{xx}$ are displayed, which show similar temperature dependencies with comparatively weak phononic maxima around 10--20\,K as is also reported for the BiSbTeSe$_2$-type materials~\cite{Sharma_THE_BSTS_2024PhRvB.109}. In order to estimate the contribution of mobile charge carriers to the heat transport we use the Wiedemann-Franz law $\kappa_{xx}^{\rm WF} = L_0\, \sigma_{xx}\, T$ with Lorenz number L$_0 \simeq 2.44\times 10^{-8}$ V$^2$K$^{-2}$, temperature $T$, and the measured $\sigma_{xx}$ of the better conducting sample S2. As shown by the open red dots at the bottom of Fig.~\ref{Fig_rho_kappa}(b), this estimate yields a tiny charge-carrier contribution at low temperatures and, despite increasing with T, it remains significantly below 10\,\% of even the lower $\kappa_{xx}$ of S1. Thus, the longitudinal heat transport of both samples is clearly phonon dominated. 
However, in contrast to the better charge-compensated BiSbTeSe$_2$-type materials~\cite{Sharma_THE_BSTS_2024PhRvB.109}, there is a small overall decrease of  $\kappa_{xx}$ as a function of increasing magnetic field.

\begin{figure}[!tbp]
	\centering
	\hspace*{-0.6cm}
	\includegraphics[width=\columnwidth]{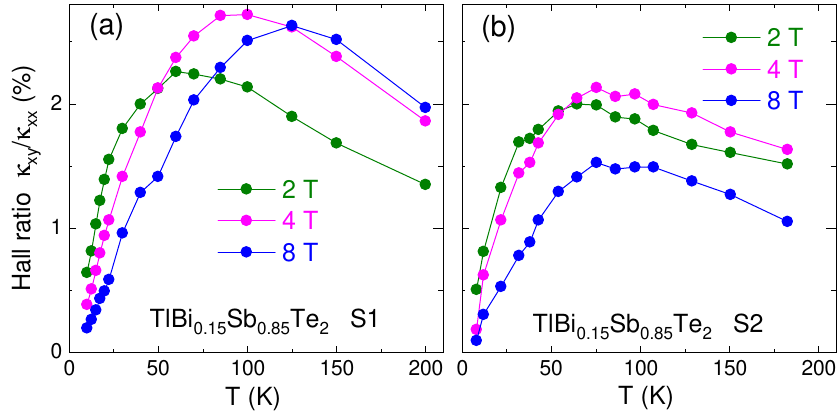}  
	\caption{Hall ratios $\kappa_{xy}/\kappa_{xx}$ of TlBi$_{0.15}$Sb$_{0.85}$Te$_2$ samples S1 and S2 are drastically enhanced to about 2\,\% at moderate magnetic fields from about 2 to $8\,$T over a wide temperature range of more than 100\,K.}
	\label{Fig_Hall_ratio}
\end{figure}

Figures~\ref{Fig_rho_kappa}(c,d) display the thermal Hall conductivities $\kappa_{xy}$ of both samples, which reveal broad maxima as a function of temperature. With increasing field, the maxima broaden even stronger and their positions move to higher temperature. As a consequence, the Hall conductivity remains large, e.g. $\kappa_{xy}\approx 50\,$mW/Km, at moderate magnetic fields from about 2 to 8\,T over a wide temperature range of about 100\,K.  
Note that the peaks in $\kappa_{xy}$ and $\kappa_{xx}$  occur at different temperatures. In contrast, for many other phonon-dominated compounds, the peaks in $\kappa_{xy}$ and $\kappa_{xx}$ occur at similar temperatures \cite{2022PNAS..11908016C,PhysRevB.105.115101,Black-Phos_Li_2023NatCo..14.1027L,Universal_phonon-THE_Jin-PRL2025,Ataei2024}. To the best of our knowledge, this is the first experimental observation of a strong THE in any material with a phonon-dominated $\kappa_{xx}$ that hardly varies with a magnetic field, whereas the position and/or shape of the $\kappa_{xy}$ peak strongly varies with field. 
Again, we estimate the contribution from mobile charge carriers to the thermal Hall conductivity by using the Wiedemann-Franz law $\kappa_{xy}^{\rm WF} = L_0\, \sigma_{xy}\,T$ with the electrical Hall conductivities measured on the same samples. As is shown by the open symbols in Fig.~\ref{Fig_rho_kappa}(c,d) this contribution is definitely negligible below about 70\,K. At higher temperature and magnetic fields it becomes sizable and can even reach up to almost 20\,\% of the measured $\kappa_{xy}$. It is, however, also clear from Fig.~\ref{Fig_rho_kappa}(c,d) that this charge-carrier contribution cannot  explain the measured $\kappa_{xy}$, because apart from being much too small at low $T$, its magnetic-field dependence is partly opposite to that of $\kappa_{xy}$ as is shown in Fig.~\ref{AppB:Fig-sigma_xy}(a,b) of Appendix~\ref{AppB:sigma}. 

\begin{figure}[!tbp]
	\centering
	\includegraphics[width=\columnwidth]{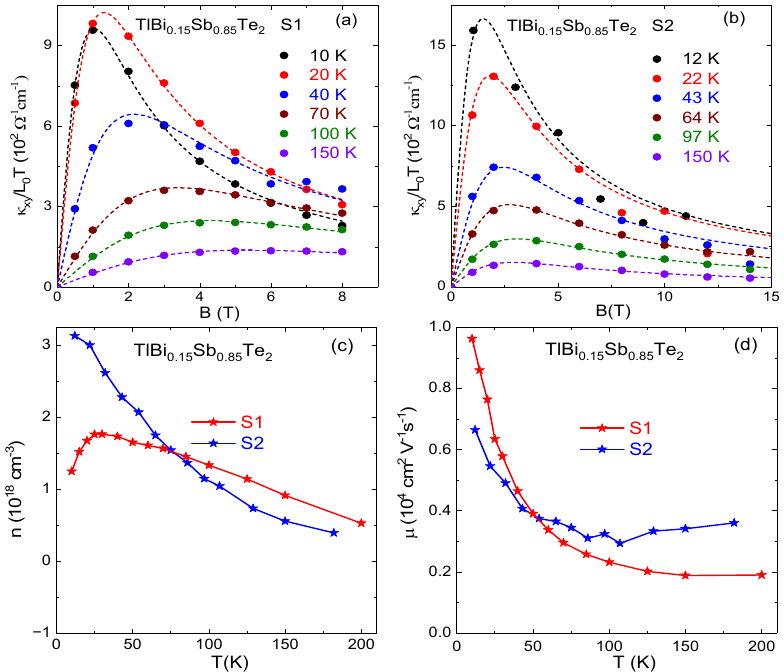} 
	\caption{ (a,b) Field dependent thermal Hall conductivity $\kappa_{xy}$ of TlBi$_{0.15}$Sb$_{0.85}$Te$_2$ samples converted to electrical Hall conductivity units by using Wiedemann-Franz law. Dashed lines are fits of a single-band model of hole-like charge carriers, see Eq.~\eqref{eq:Hall1}. (c,d) Temperature dependent fit parameters of the obtained hole-like carrier densities $n(T)$ and their mobilities $\mu(T)$ for both samples S1 and S2. }
	\label{Fig_k_xy_fits}
\end{figure}

Fig.~\ref{Fig_Hall_ratio} and Fig.~\ref{Fig_k_xy_fits}  display the main experimental result of this work. The thermal Hall ratio in TlBi$_{0.15}$Sb$_{0.85}$Te$_2$ is drastically enhanced and reaches values larger than 2\,\%. It is one of the largest thermal Hall ratios reported to date~\cite{PhysRevB.105.115101,Ataei2024}, which remarkably 
appears at moderate fields of 2 to 4\,T already, see Fig.~\ref{Fig_Hall_ratio}, and remains that large over a wide temperature range of more than 100\,K. This results from a strongly nonmonotonic magnetic-field dependence of $\kappa_{xy}(B)$ which systematically sharpens up with decreasing temperature, see Fig.~\ref{Fig_k_xy_fits}(a,b). In contrast, previous works on other insulators generally report field-linear $\kappa_{xy}(B)$ curves with temperature dependent slopes that roughly follow the temperature dependence of the phonon-dominated longitudinal heat conductivity 
$\kappa_{xx}(T)$. Thus, the largest thermal Hall effects are typically observed close to the low-temperature maxima of $\kappa_{xx}$~\cite{2022PNAS..11908016C, Black-Phos_Li_2023NatCo..14.1027L,Universal_phonon-THE_Jin-PRL2025} at highest fields.

As is shown by the dashed lines in Fig.~\ref{Fig_k_xy_fits}(a,b), the nonmonotonic magnetic-field dependencies of the thermal Hall conductivity data of both TlBi$_{0.15}$Sb$_{0.85}$Te$_2$ 
samples are very well reproduced by the  simple formula  
\begin{equation}
	\frac{\kappa_{xy}}{L_0T} = n \frac{e \mu^2 B} { 1 + \mu^2 B^2} 
	\label{eq:Hall1}
\end{equation} 
with the $T$-dependent fit parameters $n$ and $\mu$ shown in Fig.~\ref{Fig_k_xy_fits}(c,d). Here, the thermal Hall conductivity data were converted to electrical units, because Eq.~(\ref{eq:Hall1}) describes the electrical Hall conductivity $\sigma_{xy}(T,B)$ of a single-band Drude model of hole-like charge carriers with $\mu B=\omega_c \tau \ll 1\,(\gg 1)$ denoting the low-(high-)field limit. At this point, it is important to note that Eq.~(\ref{eq:Hall1}) would be naturally expected for a (homogeneous metallic) system with dominant electronic heat transport such that charge and heat conductivities are related via the Wiedemann-Franz law ($\kappa_{ij}^{\rm WF} = L \sigma_{ij}$ with $L\simeq L_0$ for dominant elastic scattering). This is not the case in these TlBi$_{0.15}$Sb$_{0.85}$Te$_2$ samples, as can be inferred already from Fig.~\ref{Fig_rho_kappa}(c,d) showing that the  Wiedemann-Franz estimated $\kappa_{xy}^{\rm WF}=\sigma_{xy} L_0 T$ is by far smaller than the measured $\kappa_{xy}$.         

Below, we will argue that $\sigma_{xy} L_0 T \ne \kappa_{xy}$ is naturally expected for an inhomogeneous system. Nevertheless, the characteristic field dependence of Eq.~(\ref{eq:Hall1}) can survive, and we provide an interpretation that the observed characteristics are clear fingerprints of a THE originating from charge puddles.

\begin{figure}
    \centering
    \includegraphics[height=.77 \linewidth]{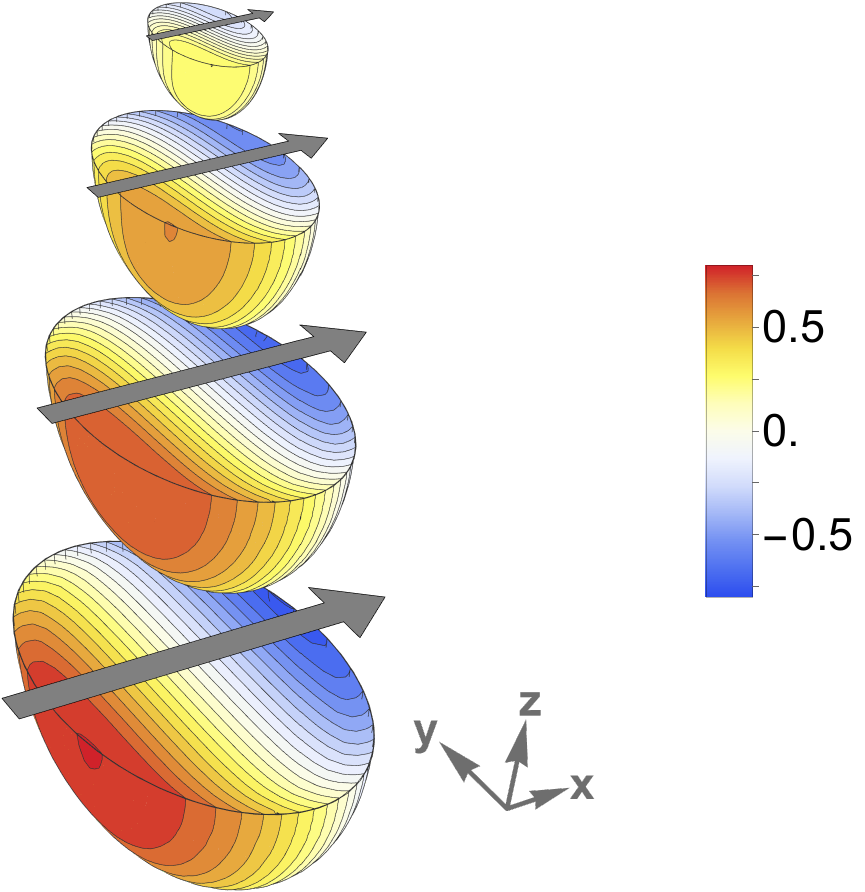}
    \caption{Heat maps of charge-carrier temperature $T^{ch}(\vec r)$ computed from Eq.~\ref{eq:heat_coupled} for   spherical  hole-doped  puddles of different sizes and a phonon heat current in the $+x$ direction (arrow). Parameters: $\mu B=2$, $R_\text{eff}=1,2,3,4$, only the lower half of the puddle, $z<0$ is shown. The Hall effect leads to an asymmetric heating and cooling.
    Larger puddles equilibrate better with the phonons. The color scale (see legend) is chosen such that $1$ and $-1$ corresponds to the phonon temperatures at the beginning and end of each puddle.
    \label{fig:heatMaps}}
\end{figure}

\section{Theory of charge puddles coupled to phonons}
In order to model the physics of charge puddles coupled to phonons, we use that in our system charge puddles are formed on length scales large compared to the distance of defects. Thus, we assume that their typical size is large compared to the electronic and phonon mean-free path for scattering. 

This allows us to describe the heat transport in a two-temperature model approximately by two coupled inhomogeneous diffusion equations,
\begin{align}
-\kappa^{ph} \vec \nabla^2 T^{ph}&= \alpha(\vec r) ( T^{ch}-T^{ph}) \label{eq:heat_coupled_ph}\\
-\vec\nabla \boldsymbol{\kappa}^{ch}(\vec r) \vec\nabla T^{ch}&= \alpha(\vec r) ( T^{ph}-T^{ch}).
\label{eq:heat_coupled}
\end{align}
For simplicity, the longitudinal phonon heat conductivity $\kappa^{ph}$ is assumed to be constant in space. In contrast, the electronic heat conductivity $\kappa^{ch}$ of electron- and/or hole-like charge carriers depends on position $\vec r$ and is finite only inside of the puddles. In the presence of an external field $B$ in the $z$ direction, $\kappa^{ch}(\vec r)$ contains a finite off-diagonal component $\kappa^{ch}_{xy}=-\kappa^{ch}_{yx}$, and the sign of $\kappa^{ch}_{xy}$ is negative (positive) for electron-doped (hole-doped) puddles, respectively. Within linear response theory, the energy transfer from the phonon to the charge-carrier system and back is proportional to the temperature difference of the two subsystems and parametrized by $\alpha(\vec r)\ge 0$, which is finite only inside of electron-(or hole-)doped puddles.

\begin{figure}[!tbp]
	\centering
	\includegraphics[width=0.8\columnwidth]{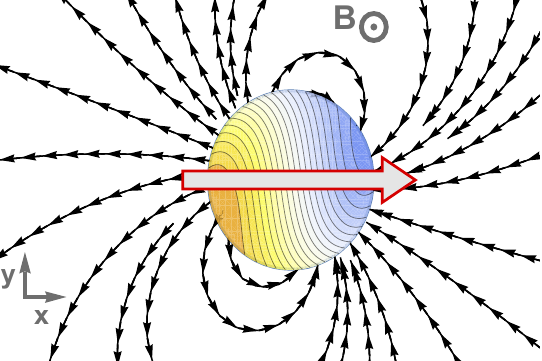} 
	\caption{Heat currents imprinted on the phonon system by a single spherical charge puddle in an infinite system.  A background phonon heat current (red arrow) induces anisotropic heating of the charge puddle indicated by the color map (parameters: $\mu B=1$, $R_{\text{eff}}=2$; see App.~\ref{app:dipole} for details; the figure shows the $z=0$ plane).
    Due to the Hall effect inside the hole-doped charge puddle, $\kappa_{xy}^{ch}>0$, the puddle gets warmer on the lower side. 
    This  temperature dipole imprints additional heat currents onto the phonon system (black arrows) 
    according to Eq.~\eqref{eq:heat_coupled_ph} in the form of a dipolar field, which is rotated relative to the $x$-axis.  }
	\label{FIG.imprint}
\end{figure}
The physics of a single, hole-doped puddle of spherical shape is illustrated in Figs.~\ref{fig:heatMaps} and~\ref{FIG.imprint}. A background temperature gradient of the phonon system  heats up the puddles inhomogeneously according to Eq.~\eqref{eq:heat_coupled}, see App.~\ref{app:dipole} for details. Due to $\kappa_{xy}^{ch}>0$ the $-y$ side of a hole-doped puddle gets warmer than the $+y$ side. This leads to an effective heating-cooling dipole imprinting heat currents into the phonon system according to Eq.~\eqref{eq:heat_coupled_ph}. The resulting heat currents in the phonon system are shown as black arrows in Fig.~\ref{FIG.imprint} for a single puddle. 

In the presence of many puddles, the individual dipolar contributions add up. Here, we use that Eq.~\eqref{eq:heat_coupled} has the form of a Poisson equation. Defining a heating-cooling polarization density 
\begin{align}
    \vec \nabla \vec p = \alpha(\vec r) \left(T^{ch}(\vec r)-T^{ph}(\vec r)\right),
\end{align}
we numerically compute in App.~\ref{app:dipole} the linear relation between $\vec p$ (averaged over a puddle) and the underlying thermal gradient of the phonon system,
\begin{align}
   \vec p = \kappa^{ph} \vec \chi \nabla_\| T^{ph}.
\end{align}
Here, $\chi$ is a dimensionless susceptibility encoding the heating and cooling of puddles. For $\kappa^{ph}\gg \kappa^{ch}$, the average transverse polarization determines the transverse thermal gradient imprinted on the phonon system, see App.~\ref{app:kappa_tot}. From the condition, that there is no net heat current in the transverse direction, one obtains
\begin{align}
   \frac{\kappa_{xy}}{\kappa_{xx}} \approx \langle \chi_\perp \rangle,\label{eq:kappa_chi}
\end{align}
where $\langle \dots \rangle$ denotes a spatial average.

\begin{figure}
    \centering
  \includegraphics[width=0.7 \linewidth]{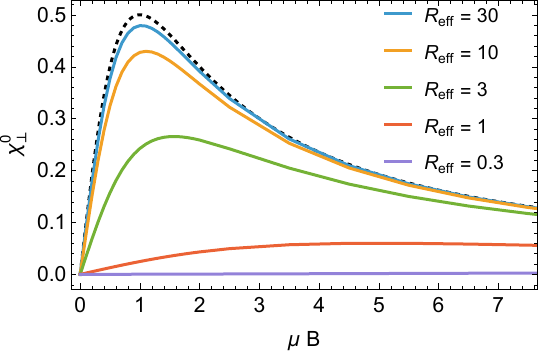}
    \caption{Field dependence of the transverse  heating susceptibility $\chi_\perp^0$ with $\chi_\perp=\chi_\perp^0  \frac{\kappa^{ch}_0}{\kappa^{ph}}$ for a spherical charge puddle with effective radius $R_\text{eff}$, computed numerically from Eq.~\eqref{eq:heat_coupled}, see  App.~\ref{app:dipole}. 
    $\chi_\perp$ determines the field-dependence of the thermal Hall effect, Eq.~\eqref{eq:kappa_chi}. The dashed line corresponds to $\mu B/(1+(\mu B)^2)$.}
    \label{fig:chiperp}
\end{figure}
In App.~\ref{app:dipole} we provide a detailed theory of the $\chi_\perp$ for spherical puddles as a function of puddle radius, coupling parameter $\alpha$, electron-phonon coupling and magnetic field. 
A central parameter of the model is the length scale $\ell_{eq}=\sqrt{\kappa_{xx}^{ch}/\alpha}$ on which the charge-carrier system equilibrates with the phonons. Thus, we measure the puddle size in units of $\ell_{eq}$ by introducing the effective puddle radius
\begin{align}
R_\text{eff}=\frac{R}{\ell_{eq}}=\frac{R \alpha^{1/2}}{{\kappa^{ch}_{xx}(B=0)}^{1/2}}.\label{eq:R_eff}
\end{align}
$R_\text{eff}$ is a strongly temperature dependent quantity as $\alpha$, $\kappa^{ch}_{xx}$ and the bare puddle radius \cite{Borgwardt_PhysRevB.93.245149} all substantially depend on $T$.
Larger puddles equilibrate more efficiently with the phonon system.
In Fig.~\ref{fig:chiperp} the resulting transverse heating susceptibility is shown for different values of the effective puddle radius  $R_\text{eff}$.  Larger puddles  have larger values of $\chi_\perp$ and have a larger volume. Both factors imply that the the spatial average $\langle \chi_\perp \rangle$, Eq.~\ref{eq:kappa_chi}, is dominated by larger $R_\text{eff}$.

The theory simplifies considerably for $R_\text{eff}\gg 1$, where an analytic solution can be obtained (furthermore, an analytical solution can be found for arbitrary $R_\text{eff}$ in the limit $\mu B\ll 1$, see App.~\ref{app:dipole}). For large $R_\text{eff}$, 
 phonon and charge-carrier temperatures equilibrate efficiently, such that $T^{ph}\approx T^{ch} \approx T$. Adding Eq.~\eqref{eq:heat_coupled_ph} and Eq.~\eqref{eq:heat_coupled}, one obtains a single equation independent of $\alpha$
\begin{align}
-\vec\nabla (\kappa^{ph} \mathbb 1+\boldsymbol{\kappa}^{ch}(\vec r))\vec\nabla T&= 0.\label{eq:heat_coupled_2}
\end{align}
Using that heat transport is dominated by phonons, leading to approximately uniform heat currents, we can employ an effective medium approximation where $\boldsymbol{\kappa}^{ch}(\vec r)$ is replaced by its spatial average. Thus, we find that  the total transverse heat conductivity is approximately given by the spatial average of the charge-carrier contribution
\begin{align}
\kappa_{xy} \approx \langle \kappa^{ch}_{xy}(\vec r)\rangle =\frac{1}{V} \int d^3 \vec r\,  \kappa^{ch}_{xy}(\vec r), \label{eq:imprint}
\end{align}
where $V$ is the volume of the system. While charge puddles  occupy only a small fraction of the volume of the system, they directly determine the THE.
This simple, central result relies on two assumptions: (i) the equilibration of the phonons and charge carriers within the puddles and (ii) a good thermal conduction due to phonons especially in the region between the puddles. These two effects ensure that transverse energy currents generated in the puddles are directly transferred to the phonons.

In contrast, charge transport is {\em not} obtained from an average of the electric conductivity $\sigma$ as it is dominated by the insulating regions between the puddles which are the bottlenecks for electrical conduction. In Refs.~\cite{PhysRevLett.121.147201,PhysRevX.8.031032}, a result similar to Eq.~\eqref{eq:imprint} was obtained for chiral spin liquids, where spin–phonon coupling  transfers the quantized thermal Hall conductivity $\kappa_{xy}$  of the edge states of a chiral spin liquid to the phonon system.

In an ideal particle-hole symmetric system $\langle \kappa^{ch}_{xy}\rangle$ vanishes by symmetry. In real materials compensation doping is not perfect and the density of electron- and hole-doped puddles will be different. Furthermore, electrons and holes will have different mobilities and different coupling constants $\alpha$. As we will discuss below, our experiment is well described by the case when  hole-doped puddles dominate  $\langle \kappa_{xy}^{ch} \rangle$.

\begin{figure}[tbp]
	\centering
	\includegraphics[width=\columnwidth]{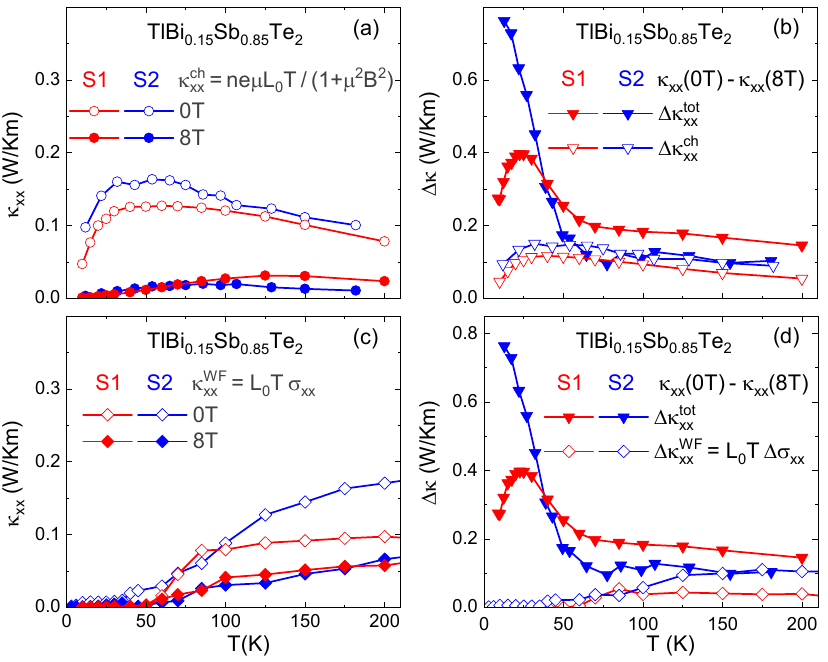}
	{\caption{(a) Estimated contribution $\kappa_{xx}^{ch}$ from mobile charge carriers to the longitudinal heat transport of samples S1 and S2 for B\,=\,0 and 8\,T, derived from the single-band fits of the measured thermal Hall conductivity $\kappa_{xy}$ data, see Fig.~\ref{Fig_k_xy_fits}. As the estimate assumes that only hole-doped puddles are present, it is only a lower bound to this contribution. (b) Comparison of the field-induced suppression of the estimated $\Delta\kappa_{xx}^{ch}(B=8\,{\rm T})$ with the measured reduction of the total heat conductivity  $\Delta\kappa_{xx}^{tot}=\kappa_{xx}^{tot}(0\,{\rm T})-\kappa_{xx}^{tot}(8\,{\rm T})$. (c) Estimates of $\kappa_{xx}^{\rm WF}$ derived from $\sigma_{xx}$ using the Wiedemann-Franz law and (d) comparison of its field-induced reduction with $\Delta\kappa_{xx}^{tot}$.}
		\label{fig:kappaxx}}
\end{figure}

\section{Interpretation}
To model the electronic heat conductivity within a single puddle $\vec \kappa^{ch}$, we use a simple relaxation time approximation, reviewed in App.~\ref{app:relax}, and obtain
\begin{align}
\frac{\kappa^{ch}_{xy}}{L_0 T}& \approx  n_i \frac{e \mu_i^2 B}{1+\mu_i^2 B^2}   ,
\end{align}
where $n_i$ is the hole density (which takes negative values for electron doping).
In the large-puddle limit, $R_\text{eff}\gg 1$, one simply averages over all puddles to obtain $\kappa_{xy}$. Experimentally, we find that a single mobility $\mu$ fits the observed field dependence remarkably well, see Fig.~\ref{Fig_k_xy_fits}. This suggests that either electron and hole puddles have similar mobilities or that one type of puddles dominates. 
The fit parameter $n$ in  Fig.~\ref{Fig_k_xy_fits}(c) can thus be interpreted as the effective density of holes in the total system which is of the order of $10^{18}\,\text{cm}^{-3}$ with some variation between our two samples. Furthermore, large mobilities  in the range of  $5000\,$cm$^{2}$V$^{-1}$s$^{-1}$ {\em within} typical puddles are obtained (while outside of the puddles transport is strongly suppressed). Importantly, these parameters are roughly in the range of values expected for charge puddles. For example, in Ref.~\cite{PhysRevB.96.195135} a total charge-carrier density of $4\times 10^{17}\,\text{cm}^{-3}$ was obtained from fits to the optical conductivity of a somewhat cleaner sample of BiSbTeSe$_2$. Furthermore in Ref.~\cite{Breunig_2017NatCom_815545B} a remarkably large puddle mobility of $1.1\times 10^{4}\,$cm$^{2}$V$^{-1}$s$^{-1}$ was fitted to describe magnetotransport in a sample of TlBi$_{0.15}$Sb$_{0.85}$Te$_2$ (with nominally the same composition as the actual S1 and S2 samples), where a percolation transition of the charge puddles was triggered upon increasing magnetic field.
Therefore the order of magnitude of the fitting parameters is consistent with our interpretation. The proximity to the percolation transition is a further strong argument in favor of large puddles, $R_\text{eff}\gg 1$, see Ref.~\cite{Breunig_2017NatCom_815545B} for a microscopic theory of the puddle structure in the material.

In contrast, thermal Hall data on a series of {\em different} topological insulators,
Bi$_{2-x}$Sb$_x$Te$_{3-y}$Se$_y$  and Bi$_{1.08}$Sn$_{0.02}$Sb$_{0.9}$Te$_2$S, reported by us in Ref.~\cite{Sharma_THE_BSTS_2024PhRvB.109}
showed an order of magnitude smaller Hall ratio of $10^{-3}$ displaying a simple linear $B$ dependence up to 15\,T. As similar types of puddles occur in these systems the question arises why $\kappa_{xy}$ behaves differently. Thus, either $R_\text{eff}\lesssim 1$ or the corresponding charge mobilities are considerably lower in these other materials. In Ref.~\cite{Borgwardt_PhysRevB.93.245149} a scattering rate of $1/\tau=140\cdot 10^{12}$/s was fitted to optical conductivity data of BiSbTeSe$_2$. Using $m^*/m=0.2$~\cite{Borgwardt_PhysRevB.93.245149}, this translates to a mobility of $\mu=e \tau/m^*=63$\,cm$^{2}$V$^{-1}$s$^{-1}$, implying a very small $\kappa_{xy}$ and a linear $B$ dependence up to $B\approx 100\,$T, consistent with the  experimental observation of Ref.~\cite{Sharma_THE_BSTS_2024PhRvB.109} even if one assumes $R_\text{eff}\gg  1$. 


Eq.~\ref{eq:heat_coupled_2} predicts that also the longitudinal conductivity $\kappa_{xx}$ obtains a field dependence arising from $\langle \kappa^{ch}_{xx} \rangle$. While in $\langle \kappa^{ch}_{xy} \rangle$ the contributions of electron- and hole-doped puddles partially cancel, they add up for $\langle \kappa^{ch}_{xx} \rangle$. The ratio of electron- and hole-doped puddles is, however, unknown and is expected to depend sensitively on both, the defect density and on temperature \cite{Borgwardt_PhysRevB.93.245149}.
Using the single-band fit of Eq.~\eqref{eq:Hall1} and assuming for simplicity that only hole-doped puddles are present, we obtain in Fig.~\ref{fig:kappaxx}(a) an estimate for the electronic contribution to $\kappa^{ch}_{xx}$, shown for $0$ and 8\,T. Assuming that the field-dependence of the measured $\kappa_{xx}^{tot}$ arises from the electronic contribution, we compare in Fig.~\ref{fig:kappaxx}(b) the measured difference $\Delta \kappa_{xx} = \kappa_{xx}(0\,\text{T})-\kappa_{xx}(8\,\text{T})$ to the difference of the estimated electronic contributions $\Delta \kappa^{ch}_{xx}$. While the order-of-magnitudes roughly fit, the measured field-induced changes are bigger, as expected for a system where both, electron- and hole-doped puddles, are present. The origin of the kink in the $T$ dependence of the measured $\Delta \kappa$ at $T\approx 50\,$K is not clear, but previous experiments \cite{Borgwardt_PhysRevB.93.245149} have shown that thermal excitations change the puddle density at this temperature scale. Well above this temperature scale there are also extra contributions to the heat transport arising from bulk electrical transport, which are obtained via the usual Wiedemann-Franz estimate $L_0 T\sigma_{xx}$ from measured $\sigma_{xx}(T,B)$ data. As shown in Fig.~\ref{fig:kappaxx}(c,d), these contributions are completely negligible below 50\,K, but become visible with a finite field-induced suppression above about 100\,K. 

\section{Conclusion}
The thermal Hall effect (THE) in insulators remains one of the central unresolved problems in the study of heat transport, owing to its widespread occurrence, unexpectedly large magnitude, and the lack of a unifying microscopic understanding. Our results establish TlBi$_{0.15}$Sb$_{0.85}$Te$_2$ as a system exhibiting an exceptionally large THE, in which the microscopic origin of the effect can be identified. 
The characteristic magnetic-field dependence of the THE -- together with previous experiments on this compound -- provides compelling evidence that charge puddles created by dilute charged impurities act as the source of the large thermal Hall response. The fact that the peak in the Hall ratio as a function of $B$ occurs roughly at $\mu B \sim 1$ using the mobility values independently determined in Ref.~\cite{Breunig_2017NatCom_815545B} provides a semi-quantitative link between the electronic properties of puddles and the macroscopic thermal Hall response.

An important open question raised by our findings is whether this mechanism is specific to a narrow class of materials or whether it reflects a more generic route to large phonon thermal Hall effects in insulating systems.
From the experimental point of view, the THE observed in  
TlBi$_{0.15}$Sb$_{0.85}$Te$_2$ is clearly {\em different} from that previously reported in insulators. It is larger than in most compounds and the characteristic field dependence has not been observed in other systems. From the theory perspective, Bi-based TI compounds are special due to their large dielectric constant, $\epsilon\approx 10^2$. As discussed in the introduction, this leads to very shallow bound states and to large well-conducting charge puddles, which are needed within our theory. Luckily, the mobilities inside the hole-doped puddles of TlBi$_{0.15}$Sb$_{0.85}$Te$_2$  turn out to be so large that a pronounced nonmonotonic magnetic-field dependence of $\kappa_{xy}(B)$ is observable, which we use as a smoking-gun signature for a puddle-induced thermal Hall effect.

It is, however, plausible that charged impurities have a strong impact on the thermal Hall effect in a much broader class of systems, where our theory cannot be directly applied. One limiting case of this problem was studied by Flebus and MacDonald \cite{PhysRevB.105.L220301}, who argued that the Lorentz force acting on charged defects in ionic crystals can give rise to Hall ratios of order $10^{-3}$. Their analysis, however, considered only the motion of the ions and did not take into account that the diverging potential of randomly placed charges necessarily creates electronic defects \cite{ShklovskiiEfros1972}, as discussed in the introduction. It therefore remains an interesting open theoretical problem to explore the scattering of phonons from such dynamical defects in regimes where they do not form well-conducting puddles. On general grounds, it has, for example, been argued by Sun, Chen, and Kivelson \cite{PhysRevB.106.144111}, as well as by Guo, Joshi, and Sachdev \cite{2022PNAS..11915141G}, that inelastic scattering from dynamical defects is a prime candidate for explaining phonon thermal Hall effects. From this perspective, scattering from large charge puddles may be viewed as one limiting case of a more general mechanism in which charged impurities generate dynamical electronic defects. Further experimental and theoretical work will be necessary to fully explore this physics.




\appendix

\section{Theory of transverse heat conductivity induced by charge puddles}\label{app:theory}
In this appendix, we investigate how the coupling of charge puddles to the phonon system  affects $\kappa_{xy}$.
Our starting point is the two-temperature model defined in  the main part of the text, Eq.~\eqref{eq:heat_coupled}, which is reproduced here.
\begin{align}
-\kappa^{ph} \vec \nabla^2 T^{ph}&= \alpha(\vec r) ( T^{ch}-T^{ph})\label{eq:heat_Tph} \\
-\vec\nabla \boldsymbol{\kappa}^{ch}(\vec r) \vec\nabla T^{ch}&= \alpha(\vec r) ( T^{ph}-T^{ch}).\label{eq:heat_coupled_app}
\end{align}
Importantly, $\alpha(\vec r)$ and $\kappa^{ch}$ are finite only within the charge puddles while $\kappa^{ph}$ is space independent. Furthermore, we assume that heat transport is dominated by phonons, $\kappa^{ph}\gg \kappa^{ch}_{xx}$.

\subsection{Puddles as heating dipoles}\label{app:dipole}

We first consider a single puddle. For simplicity, we model it as a spherical object
with a space-independent thermal conductivity
\begin{align}
    \vec \kappa^{ch} =\kappa^{ch}_0 \begin{pmatrix} \frac{1}{1+(\mu B)^2} & \frac{\mu B}{1+(\mu B)^2} & 0 \\ -\frac{\mu B}{1+(\mu B)^2} & \frac{1}{1+(\mu B)^2} & 0 \\ 0 & 0 & 1 \end{pmatrix},
\end{align}
where $\mu$ is the mobility and $\kappa^{ch}_0$ is the zero-field conductivity, see Sec.~\ref{app:relax} for a derivation within the relaxation time approximation. The sign convention is for a hole-doped puddle. We solve the diffusion equation for $T^{ch}$, Eq.~\eqref{eq:heat_coupled_app}, with 
the boundary condition that there is no electron heat current outside of the puddle. Therefore, the boundary condition at the surface of the puddle is given by \begin{align}
    \vec n \cdot \vec{\kappa^{ch}} \vec \nabla T^{ch}=0,
\end{align} where $\vec n$ is the surface normal. Note that $\kappa^{ch}_{xy}=-\kappa^{ch}_{yx}$ drops completely from the bulk diffusion equation and enters only via the boundary condition.

As an approximation, we assume that $T^{ph}$ inside the puddle varies linearly in space 
\begin{align}
T^{ph}(x) \approx T + x \nabla_\| T^{ph}.
\end{align}

Using rescaled coordinates, $\vec r \to \tilde r=\vec r/R$, where $R$ is the radius of the puddle, the diffusion equation can be written in the form
\begin{align}
-\tilde{\vec\nabla} \frac{\boldsymbol{\kappa}^{ch}}{\kappa^{ch}_0} \tilde{\vec\nabla} t^{ch}&= (R_\text{eff})^2 (\tilde x-t^{ch})\label{eq:diffe_rescaled}\\
R_\text{eff}&=\frac{ R \sqrt{\alpha}}{\sqrt{\kappa^{ch}_0}},\quad T^{ch}=T+t^{ch} \, R \nabla_\| T^{ph},
\end{align}
where the rescaled radius of the spherical puddle is $1$. We solve the rescaled diffusion equation, which depends only on two dimensionless parameters, $R_\text{eff}$ and $\mu B$, numerically (using the NDSolveValue command in Mathematica) and analytically in the small-$B$ limit, see below.


Due to $\kappa^{ch}_{xy}$, the puddle gets warmer on one side and colder on the other, see Fig.~\ref{fig:heatMaps}.
As the diffusion equation for $T^{ph}$ has the form of a Poisson equation for a charge density $\alpha(\vec r)(T^{ch}-T^{ph})$
(with dielectric constant $\kappa^{ph}$), we calculate for a puddle with radius $R$ an effective heating-cooling dipole moment defined by
\begin{align}
\vec P &=  \int d^3 r \, \alpha\, (T^{ch} -T^{ph}) \vec r=  \tilde{\vec P}\,\kappa_0^{ch} R^3 \nabla_\| T^{ph}  \nonumber \\
\tilde{\vec P} &= (R_\text{eff})^2\int d^3 \tilde r  \, (t^{ch}-\tilde x)\ \tilde{\vec{r}}.
\end{align}
The polarization density $\vec P/V$ where $V=\frac{4}{3} \pi R^3$ for a spherical puddle is linear in $\nabla_\| T^{ph}$,
\begin{align}
    \frac{\vec P}{V} &= \kappa^{ph} \vec \chi \nabla_\| T^{ph} \nonumber \\  \vec \chi &=\frac{\kappa_0^{ch}}{\kappa^{ph}}\vec \chi^0 ,\quad \vec \chi^0 =\frac{3}{4 \pi} \tilde{\vec P}, \label{eq:chi}
\end{align}
where the dimensionless `heating-cooling susceptibility' $\vec \chi$ describes the amount of polarization density is induced by $\nabla_\| T^{ph}$ the bare phonon current (the transverse gradient $\nabla_\perp T^{ph}$ will be taken into account below). Here $\vec \chi$
takes over the role of the electric susceptibility in an electrostatic problem. $\vec \chi$ is a tensor but takes the form of a vector in our equation because we used a convention where the gradient of $T^{ph}$ is in the $x$-direction.

\begin{figure}
    \centering
  \includegraphics[width=0.7 \linewidth]{chiPerp.pdf}\\[0.5cm]
   \includegraphics[width=0.7
   \linewidth]{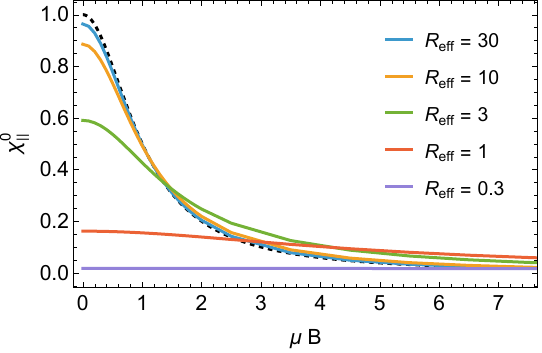}
    \caption{Field dependence of the transverse and longitudinal heating susceptibility $\chi_\perp^0$ defined in Eq.~\eqref{eq:chi}. $\chi_\perp^0$ determines the field-dependence of the thermal Hall effect. In the large-puddle limit, $R_\text{eff}\to \infty$ the analytical results of Eq.~\eqref{eq:chi_large_alpha} are obtained, dashed lines.}
    \label{fig:heatDipole}
\end{figure}
In Fig.~\ref{fig:heatDipole}
the numerically obtained values for $\chi_0$ are shown. For   $R_\text{eff}\to 0$, the heating-cooling susceptibilites are proportional to $\alpha^2$, $\chi_\perp^0\approx  (R_\text{eff})^4\, \mu B/25$ (the prefactor is calculated analytically below), and the characteristic peak at $\mu B\approx 1$ is not visible anymore.
For large $R_\text{eff}$, in contrast, we have argued in the main text of the paper, Eq.~\eqref{eq:heat_coupled_2}, that the heat condutivities of electrons and phonons are additive, which implies that 
\begin{align}
\chi^0_\perp(R_\text{eff} \to \infty)&=\frac{\kappa^{ch}_{xy}}{\kappa^{ch}_{0}}=\frac{\mu B}{1+\mu^2 B^2}
\nonumber \\
\chi^0_\|(R_\text{eff} \to \infty)&=\frac{\kappa^{ch}_{xx}}{\kappa^{ch}_{0}}=\frac{1}{1+\mu^2 B^2}.\label{eq:chi_large_alpha}
\end{align}
The dashed lines in Fig.~\ref{fig:heatDipole} confirm this analytical result, see Sec.~\ref{app:relax}.

An analytic solution can furthermore be obtained in the small $B$ limit for arbitary values of $R_\text{eff}$. In this case, we write (in rescaled coordinates) $t^{ch}(\tilde{\vec r})=t^{ch}_0(\tilde{\vec r})+t^{ch}_1(\tilde{\vec r})$, where $t^{ch}_0$ is the change of the (rescaled) charge-carrier temperature for $B=0$ while $t^{ch}_1$ is linear in $B$. To obtain $t^{ch}_0$ and $t^{ch}_1$, we use the ansatz 
\begin{align}
t^{ch}_0&=\tilde x f_0(\tilde r), \qquad t^{ch}_1=\tilde y f_1(\tilde r),
\end{align} where $\tilde r=|\tilde{\vec r}|$. By solving the resulting radial DGL together with the boundary condition at $\tilde r=1$, we obtain
\begin{align}
   f_0(\tilde r)&=1+c_0  \frac{\tanh(\tilde{r}R_\text{eff}) - \tilde{r}R_\text{eff}}{\tilde{r}^3 }\,
   \nonumber \\
   f_1(\tilde r)&=b \, c_1 \,\frac{\tanh(\tilde rR_\text{eff}) - \tilde rR_\text{eff}}{\tilde r^3}.
\end{align}
with $b=\kappa^{ch}_{xy}/\kappa^{ch}_0=\mu B$ evaluated for $\mu B\ll 1$ and the constants $c_0= \frac{1}{ (R_\text{eff}^2 + 2)\tanh(R_\text{eff}) - 2R_\text{eff} }$ and $c_1=\,
    \frac{(R_\text{eff}^2+3)\tanh(R_\text{eff}) - 3R_\text{eff} }{\left[(R_\text{eff}^2+2)\tanh(R_\text{eff}) - 2R_\text{eff} \right]^2}$. The argument $\tilde r R_\text{eff}=\frac{r}{\ell_{eq}}$ is the ratio of the distance $r$ from the center of the puddle and the
    thermalization length $\ell_{eq}$. The $\tilde r$ dependence of the two formulas is identical as $\kappa^{ch}_{xy}$ enters only in the boundary condition of our problem.
Using $t^{ch}_1$, we compute $\chi_\perp^0$ and obtain
for $\mu B \ll 1$ the analytic result valid for arbitary $R_\text{eff}$
\begin{align}
\chi_\perp^0(\mu B \ll 1) = & b \left( 1 + \frac{\tanh(R_\text{eff}) - R_\text{eff}}{(R_\text{eff}^2 + 2)\tanh(R_\text{eff}) - 2R_\text{eff}} \right)^2 \nonumber \\
\approx & \, \mu B \times \left\{\begin{array}{ll}
1 & \text{for } R_\text{eff} \gg 1\\[1mm] 
\frac{1}{25} R_\text{eff}^4=\frac{\alpha^2 R^4}{25 (\kappa_0^{ch})^2}  & \text{for } R_\text{eff} \ll 1
\end{array}\right. \label{eq:chiperpana}.
\end{align}
Note that in the perturbative small-puddle limit, $\chi_\perp^0$ is quadratic in the electron-phonon scattering rate $\alpha$. 

\subsection{Total heat conductivity of the phonon system}\label{app:kappa_tot}
In Sec.~\ref{app:dipole}, we solved diffusion equation for $T^{ch}$ for a single puddle. As pointed out above, we can use the fact that the heat diffusion equation for $T^{ch}$, Eq.~\ref{eq:heat_Tph}, takes the form of the Poisson equation of electrostatics, which can be written as
\begin{align}
-\vec \nabla^2 T^{ph}&= \frac{1}{\kappa^{ph}} \vec \nabla \vec p,
\end{align}
where $\vec p$ is the heating-cooling polarization density obtained from 
$\nabla \vec p=\alpha (T^{ch}-T^{ph})$. Each puddle acts as a dipole. The transverse dipoles with $p_\perp=\chi_\perp \nabla_\| T^{ph}$ sum up to create a net temperature gradient perpendicular to the heat current.

The analogy of the Poisson equation and the diffusion equation is not perfect as the boundary condition of vanishing energy current corresponds to a vanishing electric field, $\vec n \vec E=0$, where $\vec n$ is the surface-normal, which is not a common boundary condition in electrostatics. In magnetostatics, in contrast, such a boundary condition can be realized by a superconductor. In this case $\vec p$ corresponds to the magnetization.

To compute the Hall ratio $\frac{\kappa_{xy}}{\kappa_{xx}}=\frac{\nabla_y T^{ph}}{\nabla_x T^{ph}}=\frac{\nabla_\perp T^{ph}}{\nabla_\| T^{ph}}$, we assume that the phonon heat conductivity dominates in the system. In this case, $\vec \chi \ll 1$ and we can ignore the difference of $\nabla T^{ph}$ inside and outside of the puddle (otherwise, one has to take into account an extra correction to the effective medium approximation as reviewed and checked numerically, e.g., in Ref.~\cite{RevModPhys.45.574}). Furthermore, we can simply average over the polarization. From the condition of a vanishing surface energy current $\vec n (\vec \nabla T^{ph}+\frac{1}{\kappa^{ph}} \vec p )= 0 $, we obtain 
\begin{align}
\frac{\kappa_{xy}}{\kappa_{xx}}\approx \langle \chi_\perp \rangle = \frac{\kappa^{ch}_0}{\kappa^{ph}}\langle \chi^0_\perp\rangle, 
\end{align}
where $\langle \dots \rangle$ describes a spatial average.  Thus, the field dependence of $\chi^0_\perp$ shown in Fig.~\ref{fig:heatDipole} determines directly the field dependence of the Hall ratio.

For $R_\text{eff} \gg 1$ and assuming that only one type of puddle dominates, we thus obtain 
\begin{align}
\frac{\kappa_{xy}}{\kappa_{xx}} \approx \frac{\langle \kappa^{ch}_{xy}\rangle}{\kappa^{ch}_{0}}=\frac{\langle \kappa^{ch}_0 \rangle}{\kappa^{ph}} \frac{\mu B}{1+\mu^2 B^2}\approx \frac{\langle \kappa^{ch}_{xy}\rangle}{\kappa^{ph}},
\end{align}
where the spatial average $\langle \kappa^{ch}_{xy}\rangle$ is proportional to the volume fraction of puddles.
For $R_\text{eff} \ll 1$, instead, one finds
\begin{align}
\frac{\kappa_{xy}}{\kappa_{xx}} \approx \frac{1}{25} \frac{\langle \kappa^{ch}_0  R_\text{eff}^4\rangle }{\kappa^{ph}} \,\mu B\approx \frac{1}{25} \left\langle\frac{\alpha^2 R^4 (\vec \kappa_{ch}^{-1})_{xy}}{\kappa^{ph} }\right\rangle,
\end{align}
where the prefactor is obtained from Eq.~\eqref{eq:chiperpana}.
In this limit, the Hall angle is determined by the average thermal Hall resistivity of the electronic system which is linear in $B$ and independent of $\mu$ in our approximation, see Eq.~\eqref{eq:rho}. 

\subsection{Heat transport in relaxation time approximation}\label{app:relax}
Here, we briefly review the well-known theory of heat conductivity of electrons or holes in relaxation time approximation. 

The Boltzmann equation for electrons in relaxation time approximation is given by
\begin{align}
(-e \vec E \vec v_k -\frac{\vec  \nabla T}{T} \vec v_k \epsilon_k) \frac{df_0}{d \epsilon_k} - e (\vec v_k \times \vec B) \frac{d f_k}{d \vec k}=-\frac{\delta f_k}{\tau},
\end{align}
where $\epsilon_k=k^2/(2 m)-\mu$ for a weakly doped semiconductor.  The Boltzmann equation is solved by the ansatz
\begin{align}
\delta f_k=- \tau \frac{df_0}{d \epsilon_k} \vec v_k \cdot \vec X_{\epsilon_k},
\end{align}
where $\vec v_k$ is the only non-rotationally-invariant part. Thus only its derivate contributes to the $df/d\vec k$ part and 
\begin{align}
(-e \vec E  -\frac{\vec  \nabla T}{T} \epsilon_k)\cdot \vec v_k=(\vec X_{\epsilon_k} + \frac{e \tau}{m} (\vec B \times \vec X_{\epsilon_k}))\cdot \vec v_k .
\end{align}
Using $\omega_c \tau= e B \tau/m$ for a field in $\hat z$ direction we obtain
\begin{align}
X_{\epsilon_k} =&\frac{1}{1+(\omega_c \tau)^2}  \\ 
&\times \left(\begin{array}{ccc}
1 & \omega_c \tau &0 \\
-\omega_c \tau  & 1 &0 \\
0 & 0 & 1+(\omega_c \tau)^2 \\
\end{array} \right) (-e \vec E  -\frac{\vec  \nabla T}{T} \epsilon_k). \nonumber\label{XM}
\end{align}

Let us define $\vec j=(j_c,j_e)$, the charge and energy currents and 
\begin{align}
\vec j =\vec L \cdot (\vec E,-\vec \nabla T/T) ,
\end{align}
where each entry in the $2 \times 2$ matrix $\vec L$ is actually a matrix. Within our approximation, it follows from Eq.~\eqref{XM} that these matrices are proportional to each other  and thus commute with each other.
Heat transport occurs under the boundary condition that no charge is flowing, $\vec j=(0,j_e)$ and thus $-\vec \nabla T/T= (L^{-1})_{22}\, j_e$ or
\begin{align}
   { \vec \kappa^{ch}} = -\frac{1}{T} ((L^{-1})_{22})^{-1}= -\frac{1}{T} (L_{22}-L_{21} L_{11}^{-1} L_{12})
\end{align}
with commuting matrices as discussed above.

As the density of electrons in the puddles can be very small, the temperature can be larger or smaller than the relevant Fermi energy. To compute transport, we evaluate the electric current, 
$\vec j_c=2 \int \frac{d^3 k}{(2 \pi)^3} \vec v_{\vec k} \delta f_{\vec k}$, and the heat current, $\vec j_e=2 \int \frac{d^3 k}{(2 \pi)^3} \vec v_{\vec k} (\epsilon_\vec{k}-\mu)\delta f_{\vec k}$. Thus
we need the following integrals.
\begin{align}
I_1&=-\int \frac{d^3 k}{(2 \pi)^3} (\vec v_k^x)^2 \frac{df_0}{d \epsilon_k}=\frac{n}{m}  \\
I_2&=-\int \frac{d^3 k}{(2 \pi)^3} \epsilon_k (\vec v_k^x)^2  \frac{df_0}{d \epsilon_k}\\
I_3&=-\int \frac{d^3 k}{(2 \pi)^3} \epsilon_k^2 (\vec v_k^x)^2  \frac{df_0}{d \epsilon_k}.
\end{align}
We can express the resulting thermal conductivity using the Lorenz number
\begin{align}
    L(\tilde \mu,T)=\frac{k_B^2}{e^2} \, \frac{I_3-I_2^2/I_1}{(k_B T)^2 I_1} .
\end{align}
where $\tilde \mu$ is the chemical potential measured from the band bottom. This quantity has a very weak $T$ dependence. $L(\tilde \mu,T\ll \tilde \mu)=L_0=\frac{k_B^2}{e^2} \frac{\pi^2}{3}$ in the quantum limit while within our relaxation time approximation the large $T$ limit,  $L(\tilde \mu,T\gg \tilde \mu)=\frac{k_B^2}{e^2} \frac{5}{2}$, is only 20\% smaller. Therefore, the $T$-dependence of the Lorenz number is of little relevance for the experiment and can be absorbed in a small correction to $n(T)$. Therefore, we approximate $L(\tilde \mu,T)\approx L_0$ in our analysis.

Combining all results and switching from electrons to holes to be consistent with the sign conventions in our plots, we find
\begin{align}
\frac{\kappa^{ch}_{xy}}{L(\tilde\mu,T) T}& =n \frac{e \mu^2 B}{1+\mu^2 B^2}   \nonumber \\
\frac{\kappa^{ch}_{xx}}{L(\tilde\mu,T) T}&=\frac{e \mu n}{1+ \mu^2 B^2}, \label{eq:kappaxyNaive}
\end{align}
where $\mu=e \tau/m$ is the mobility, $\omega_c \tau =\mu B$ and $\sigma_0=n e^2\tau/m = e \mu n$ is the electric conductivity at $B=0$ and $n$ is the hole density. 

Note that the thermal Hall resistivity, obtained from the inverse of the matrix $\vec \kappa^e$, is within the relaxation time  approximation linear in $B$ and independent of the mobility $\mu$
\begin{align}\label{eq:rho}
(\vec \kappa_e^{-1})_{xy}=\frac{ \kappa^e_{xy}}{(\kappa_e^{xx})^2+(\kappa_e^{xy})^2}
&=\frac{B}{ n\, e\, T\, L(\tilde\mu,T)}.
\end{align}


\section{Electrical Hall conductivity} 
\label{AppB:sigma}

\begin{figure}[!tbp]
	\centering
	\includegraphics[width=\columnwidth]{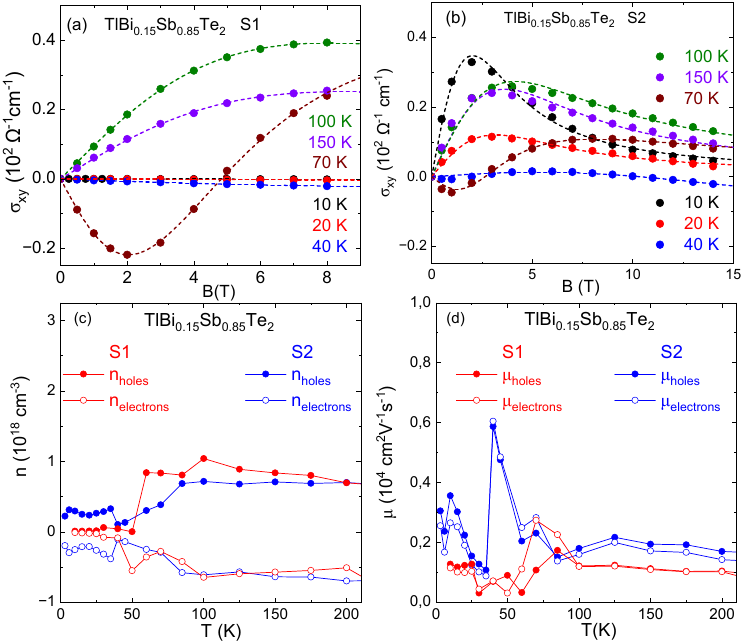} 
	\caption{ (a,b) Electric Hall conductivity $\sigma_{xy}(B,T)$ of the  S1 and S2 samples also used in the thermal Hall effect studies. The sign changes of $\sigma_{xy}(B,T)$ below 100\,K signal two-band behavior of hole- and electron-like charge carriers. The dashed lines are fits of $\sigma_{xy}$ obtained from a two-band-model analysis 
    with temperature dependent fit parameters for the hole- ($\bullet$) and electron-like ($\circ$) carrier densities $\pm n_i(T)$ and their mobilities $\mu_i(T)$ displayed in (c,d). The fit parameters were obtained by simultaneous fits of the $\sigma_{xy}(B,T)$ and $\sigma_{xx}(B,T)$ data by Eqs.~(\ref{AppB:eq-sigma_xy}, \ref{AppB:eq-sigma_xx}).
	\label{AppB:Fig-sigma_xy}}
\end{figure}

In this appendix we briefly discuss the temperature and magnetic-field dependent electrical conductivities $\sigma_{xy}(B,T)$ and $\sigma_{xx}(B,T)$ measured on the same TlBi$_{0.15}$Sb$_{0.85}$Te$_2$ samples that were used for the THE measurements. The electric Hall conductivities are plotted in Fig.~\ref{AppB:Fig-sigma_xy}(a,b) and are in fact drastically different from the corresponding thermal Hall conductivities shown in Fig.~\ref{Fig_k_xy_fits}(a,b). First of all, the low-temperature absolute values of $\sigma_{xy}$ are at least one order of magnitude smaller than the corresponding  $\kappa_{xy}/(L_0T)$ data, and, secondly, $\sigma_{xy}(B,T)$ of both samples exhibit magnetic-field and/or temperature dependent sign changes. These sign changes are clear signatures of a two-band behavior with electron- and hole-like charge carriers. Thus, two-band model fits via Eq.~\ref{AppB:eq-sigma_xy} were used to reproduce the experimental $\sigma_{xy}(B,T)$ data of both samples, which are shown by the dashed lines in Fig.~\ref{AppB:Fig-sigma_xy}(a,b). 

\begin{eqnarray}
	\sigma_{xy} = \frac{n_h e \mu_h^2 B} {1 + \mu_h^2 B^2} + \frac{n_e e \mu_e^2 B}{1 + \mu_e^2 B^2}\, 
	\label{AppB:eq-sigma_xy} \\
    \sigma_{xx} = \frac{n_h e \mu_h}{1 + \mu_h^2 B^2} - \frac{n_e e \mu_e}{1 + \mu_e^2 B^2} ,\label{AppB:eq-sigma_xx}
\end{eqnarray}
Here (and in Fig.~\ref{AppB:Fig-sigma_xy}), we use a convention with $n_h>0$ and $n_e<0$, such that electron and hole contributions are additive for $\sigma_{xx}$ but subtract for $\sigma_{xy}$.
As is frequently discussed, see e.g.\cite{two-band_model_PhysRevB.95.115126,two-band_model_TIs_PhysRevB.99.165128}, such two-band-model analyzes can be problematic, in particular, when the underlying $\sigma_{xy}(B)$ data do not show pronounced extrema and/or sign changes. In such cases, a simultaneous fit of the corresponding $\sigma_{xx}(B)$ data via Eq.~(\ref{AppB:eq-sigma_xx}) can help to stabilize the fit parameters. Such simultaneous fits yield satisfactory descriptions of $\sigma_{xy}(B)$ and $\sigma_{xx}(B)$ in the higher temperature range for both samples S1 and S2. In the lower temperature range, however, where the zero-field resistance changes to semiconductor- or insulator-like behavior, the $\sigma_{xx}$ and/or $\sigma_{xy}$ data partly show more complicated $B$ dependencies, which cannot be captured by the simplified 2-band Drude-model of Eqs.~(\ref{AppB:eq-sigma_xy}, \ref{AppB:eq-sigma_xx}). Because this mainly concerns the magneto-conductance, which according to Eq.~(\ref{AppB:eq-sigma_xx}) monotonically decreases with increasing $|B|$, we have enhanced the relative weights in the simultaneous fitting procedure towards the Hall data with the aim that, on the one hand, Eq.~(\ref{AppB:eq-sigma_xy}) still reproduces the field dependent $\sigma_{xy}(B)$, whereas, on the other hand, Eq.~(\ref{AppB:eq-sigma_xy}) remains at least close the average value of $\sigma_{xx}(B)$. The complete data sets and fits are available as supplementary material on Zenodo~\cite{zenodo_2026_18482807}.

The obtained densities and mobilities are displayed as (filled) open symbols for the (electron-) hole-like carriers in Fig.~\ref{AppB:Fig-sigma_xy}(c,d). For both samples, the carrier densities are slightly below $10^{18}\,$cm$^{-3}$ and remain almost constant above about 100~K, but decrease at lower temperature. For each sample, the mobilities of electron- and hole-like carriers show an almost identical temperature dependence, and are systematically lower in the more insulating sample S1 than in S2, see also Fig.~\ref{Fig_rho_kappa}(a). Although the parameters obtained from this two-band-model analysis cannot be taken literally, the general electrical transport behavior of these charge-compensated TIs is reasonably reproduced: at higher temperature the total conductivity is given by the sum of mobile holes and electrons, which are both of comparatively low concentration, but the holes are dominant, resulting in a positive Hall conductivity. At lower temperature, the localization of both types of charge carriers enhances the overall resistivity and causes a complex temperature and magnetic-field dependence of $\sigma_{xy}(B,T)$. In this respect the electrical transport data of our samples are in line with previous results obtained on the same and related types of TI materials~\cite{Breunig_2017NatCom_815545B,PhysRevB.84.165311,Borgwardt_PhysRevB.93.245149}. The basic new finding concerns the huge thermal Hall conductivity in both samples, which remains essentially unaffected by the localization of charge carriers and can be traced to the presence of charge puddles thermally coupled to the phonon system.

\setcounter{figure}{0}
\renewcommand{\thefigure}{S\arabic{figure}}

\acknowledgements
We acknowledge useful discussions with 
L.\ Balents, K.\ Behnia, O.\ Breunig, G.\ Grissonnanche, M.\ Gr\"uninger, L.\ Savary,  L.\ Taillefer,  and A.\ Taskin. 
This work was supported by the Deutsche Forschungsgemeinschaft (DFG, German Research Foundation) within CRC 1238 (Project-ID 277146847, subprojects A04, B01, C02).

\section*{Data availability}
The data and Mathematica code used in this study are available openly available on Zenodo~\cite{zenodo_2026_18482807}.


\providecommand{\noopsort}[1]{}\providecommand{\singleletter}[1]{#1}%

\end{document}